\documentclass[11pt]{article}

\usepackage[final]{acl}
\usepackage{float}
\usepackage[final]{acl}

\usepackage{times}
\usepackage{latexsym}

\usepackage[T1]{fontenc}
\usepackage[utf8]{inputenc}

\usepackage{microtype}

\usepackage{inconsolata}

\usepackage{graphicx}

\title{Domain-Partitioned Hybrid RAG for Legal Reasoning: Toward Modular and Explainable Legal AI for India}

\author{
\textbf{Rakshita Goel}\textsuperscript{1},
\textbf{S Pranav Kumar}\textsuperscript{1},
\textbf{Anmol Agrawal}\textsuperscript{1} \\
\textbf{Divyan Poddar}\textsuperscript{1},
\textbf{Pratik Narang}\textsuperscript{1},
\textbf{Dhruv Kumar}\textsuperscript{1} \\
\textsuperscript{1}Birla Institute of Technology and Science, Pilani \\
\textbf{Correspondence:} 
\href{mailto:f20221108@pilani.bits-pilani.ac.in}
{\texttt{f20221108@pilani.bits-pilani.ac.in}}
}

\begin{document}
\maketitle
\makeatletter
\renewenvironment{abstract}{%
    \if@twocolumn
      \section*{\abstractname}%
    \else
      \small
      \begin{center}%
        {\bfseries \abstractname\vspace{-.5em}\vspace{\z@}}%
      \end{center}%
      \quote
    \fi
}{%
    \if@twocolumn\else\endquote\fi
}
\makeatother

\begin{abstract}
India’s legal ecosystem, with over 4.55 million pending cases at the district level, requires scalable AI systems that improve legal research and access to authoritative information. Existing platforms such as SCC Online and Manupatra rely largely on keyword retrieval, often yielding weak contextual grounding. To address this, we propose a domain-partitioned hybrid RAG–KG framework tailored to the structure of Indian law.

The system integrates three specialized RAG modules covering Supreme Court case law, statutory and constitutional texts, and the Indian Penal Code each supported by optimized vector indexes for long-form legal retrieval. To capture relational knowledge beyond text similarity, we construct a large Neo4j-based Legal Knowledge Graph. An LLM-driven query router orchestrates these components by identifying the query’s domain, selecting relevant RAG modules, triggering KG traversal when relational evidence is required, and synthesizing grounded, citation-aware answers.

Evaluation using an LLM-as-a-Judge framework shows that the hybrid system achieves a 70\,\% pass rate (6.09/10), substantially outperforming a RAG-only baseline at 37.5\,\% (4.16/10). Overall, the framework provides a scalable and interpretable legal AI assistant that strengthens accuracy, grounding, and cross-domain reasoning for Indian legal research.
\end{abstract}

\section{Introduction}

Legal research in India remains highly manual and fragmented despite the availability of digital repositories and commercial legal research platforms such as LexisNexis, Westlaw, and emerging AI-assisted tools \citep{lawproducts2024a,lawproducts2024b,harvey2024}. Lawyers, judges, and scholars must routinely navigate statutes, penal provisions, and judicial precedents spread across heterogeneous sources, relying primarily on keyword-based search and manual citation tracing over constitutional texts, central acts, penal codes, and historical case records \citep{constitution,centralacts,ipcdataset,pastcases}. The dense interconnections among legal documents, including cross-referenced provisions, layered amendments, and hierarchical precedent structures, make it difficult to efficiently retrieve and synthesize relevant information using traditional retrieval pipelines \citep{reuter2025,lexgraph2025}. This gap between the availability of legal data and the accessibility of structured legal reasoning continues to hinder research efficiency, underscoring the need for intelligent, structured, and interpretable legal information systems grounded in authoritative legal sources \citep{metrics2024b,agenticlaw2025}.

Large language models (LLMs) and retrieval-augmented generation (RAG) frameworks have emerged as promising solutions by grounding generative outputs in authoritative documents. Surveys such as \citet{metrics2024b} highlight persistent challenges in retrieval noise, chunk mismatch, and multi-hop reasoning. These issues are magnified in legal settings, where documents are long and structurally complex. Prior work including \citet{reuter2025}, \citet{lexrag2025}, \citet{legalbench2024}, and \citet{lexgraph2025} shows that monolithic or purely embedding-based retrieval pipelines often struggle with citation drift and deep reasoning over statutes and case law.

Within the Indian legal domain, recent efforts have demonstrated both progress and persistent limitations. Domain-adapted models such as \citet{inlegalllama2024}, \citet{inlegalbert2022}, and the IIT Bombay study on pre-trained legal models \citep{iit2022} show substantial gains over generic LLMs. Indian-law-specific datasets including \citet{indiclegalqa2025}, \citet{nyayaanumana2024}, \citet{ailqa2024}, \citep{il_pcsr2025}, and \citep{il_tur2024} offer valuable benchmarks for QA, statute retrieval, and legal reasoning. Complementary KG-based work, such as \citet{text2kg2022}, underscores the importance of explicit relational modeling, while ranking-based approaches such as \citet{casrank2023} highlight the role of citation-aware statute retrieval.

These studies collectively emphasize two gaps: (1) the lack of unified systems that combine semantic retrieval with structured relational reasoning, and (2) the absence of agentic controllers capable of dynamically orchestrating retrieval across heterogeneous legal sources. Motivated by these gaps, we present a domain-partitioned hybrid RAG–KG architecture purpose-built for Indian law. The system integrates three domain-specialized semantic retrieval modules with a large-scale legal knowledge graph and an LLM-powered agentic orchestrator for dynamic query routing, multi-hop reasoning, and grounded synthesis. By unifying semantic retrieval with structured legal knowledge, the proposed system addresses fundamental limitations in interpretability, cross-domain reasoning, and retrieval fidelity, advancing the development of explainable and scalable legal AI for the Indian judicial context.

Evaluation using LLM-as-a-Judge framework shows that the hybrid system achieves a 70\,\% pass rate (6.09/10), substantially outperforming a RAG-only baseline at 37.5\,\% (4.16/10). Overall, the framework provides a scalable and interpretable legal AI assistant that strengthens accuracy, grounding, and cross-domain reasoning for Indian legal research.

\section{Related Work}

\subsection{Retrieval-Augmented Generation and Legal LLMs}

Retrieval-Augmented Generation (RAG) has become a prominent paradigm for reducing hallucinations through grounding in retrieved evidence, as shown by \citet{metrics2024b}. Legal-specific RAG frameworks such as LexRAG \citep{lexrag2025} and LegalBench \citep{legalbench2024} demonstrate that naïve retrieval pipelines struggle with multi-hop statutory reasoning, citation chaining, and cross-document consistency. Further, \citet{reuter2025} highlight Document-Level Retrieval Mismatch (DRM) as a core limitation when applying RAG to long, structured legal texts.

Agentic control has emerged as a promising direction for improving retrieval robustness. \citet{agenticlaw2025} demonstrate that dynamic routing across retrieval modules significantly enhances reasoning depth.

Indian research increasingly focuses on domain-specific LLMs. Models such as InLegalBERT \citep{inlegalbert2022} and InLegalLLaMA \citep{inlegalllama2024} show that pretraining on Indian statutes and case law improves downstream performance. The IIT Bombay study on pre-trained legal models \citep{iit2022} further confirms that Indian-law-specific corpora are essential for jurisdiction-aligned reasoning. However, these models still lack explicit structures for linking statutes, penal provisions, and precedents, motivating hybrid retrieval architectures such as ours.

Together, these studies indicate that while RAG-based architectures outperform naïve generative methods, they remain constrained by monolithic retrieval pipelines. Our work extends this literature by partitioning retrieval across three legal domains case law, statutory and constitutional texts, and penal codes while incorporating a knowledge graph layer to enable inter-domain reasoning.

\subsection{Knowledge Graphs and Graph-Augmented Retrieval}

Structured reasoning has been explored through graph-augmented architectures. KG-LegalRAG \citep{kglrag2025} and LexGraph \citep{lexgraph2025} represent statutes, judgments, and citations as graph nodes, enabling relational inference and reducing retrieval drift.

Indian-specific graph construction work, such as \citet{text2kg2022}, demonstrates that entity-centric relationships: articles, cases, judges, and citations significantly improve interpretability for Indian judicial texts. These findings align with global graph-based efforts (e.g., KG-LegalRAG and LexGraph), reinforcing the need to integrate a structured KG with semantic retrieval.

Our system extends these efforts by combining a 2,586-node Neo4j Legal Knowledge Graph with domain-partitioned RAG modules and an agentic orchestrator, enabling multi-hop reasoning across precedents, statutes, and IPC sections.

Our work builds on these ideas but contributes two key innovations:  

(1) an Indian-specific legal knowledge graph linking the Constitution, the Indian Penal Code, and case precedents

(2) an agentic controller that dynamically selects between RAG retrieval and graph-based inference.

\subsection{Indian Legal Benchmarks and Datasets}

India has recently witnessed a surge in the development of high-quality legal datasets enabling rigorous evaluation of language models. Notable among them is IndicLegalQA \citep{indiclegalqa2025}, which provides domain-specific QA pairs across statutory and case-law contexts. IL-TUR \citep{il_tur2024} extends this by focusing on legal text understanding and reasoning. Datasets such as NYAYAANUMANA \citep{nyayaanumana2024} and the InLegalLLaMA judgment-prediction corpus further explore predictive reasoning over judicial decision-making. Additionally, AILQA \citep{ailqa2024} offers a structured evaluation suite for legal QA, while IL-PCSR \citep{il_pcsr2025} introduces a combined statute–case retrieval benchmark for multi-source legal research.

A large body of datasets released by Saptarshi Ghosh and collaborators \citep{saptarshi2022} continues to serve as foundational resources for Indian legal NLP, spanning case retrieval, summarization, and citation prediction.
Across these datasets, recurring challenges emerge: extremely long documents, deeply nested citations, and cross-domain dependencies involving statutes, amendments, and precedents. Evaluations consistently show that models trained solely on unstructured text struggle with multi-hop reasoning and relational inference. Motivated by these limitations, we developed a 40-question synthetic benchmark inspired by IndicLegalQA, IL-TUR, NYAYAANUMANA, and AILQA, specifically targeting cross-domain reasoning patterns frequently encountered in Indian legal research.

\subsection{Commercial and Applied Legal AI Systems}

Commercial systems such as Lexis Plus AI \citep{lawproducts2024a} and 
Westlaw Edge \citep{lawproducts2024b} , and Harvey AI \citep{harvey2024} demonstrate the practical utility of AI-assisted legal research. However, these systems are proprietary, opaque, and optimized primarily for Western jurisdictions, limiting their applicability to the Indian legal context. Their limitations highlight the need for open, transparent, modular architectures such as our hybrid RAG–KG framework.

\section{Methodology}
The proposed system adopts a modular, hybrid architecture that integrates multiple retrieval-augmented generation (RAG) components with a structured legal knowledge graph (KG) for domain-specific reasoning. 
The overall methodology has been designed to address two key challenges observed in prior legal AI literature: 
(i) the inability of monolithic retrieval pipelines to adapt to heterogeneous legal document structures \cite{lexrag2025,legalbench2024}, and 
(ii) the lack of explicit reasoning links between statutes, precedents, and penal provisions that often lead to retrieval drift or incomplete factual grounding \cite{lexgraph2025,agenticlaw2025}. 
By combining specialized RAGs, a relational graph, and a planned agentic workflow, our approach seeks to achieve accurate retrieval, contextual synthesis, and interpretable reasoning across diverse Indian legal sources.

\begin{figure}[h!]
    \centering
    \includegraphics[width=0.9\linewidth]{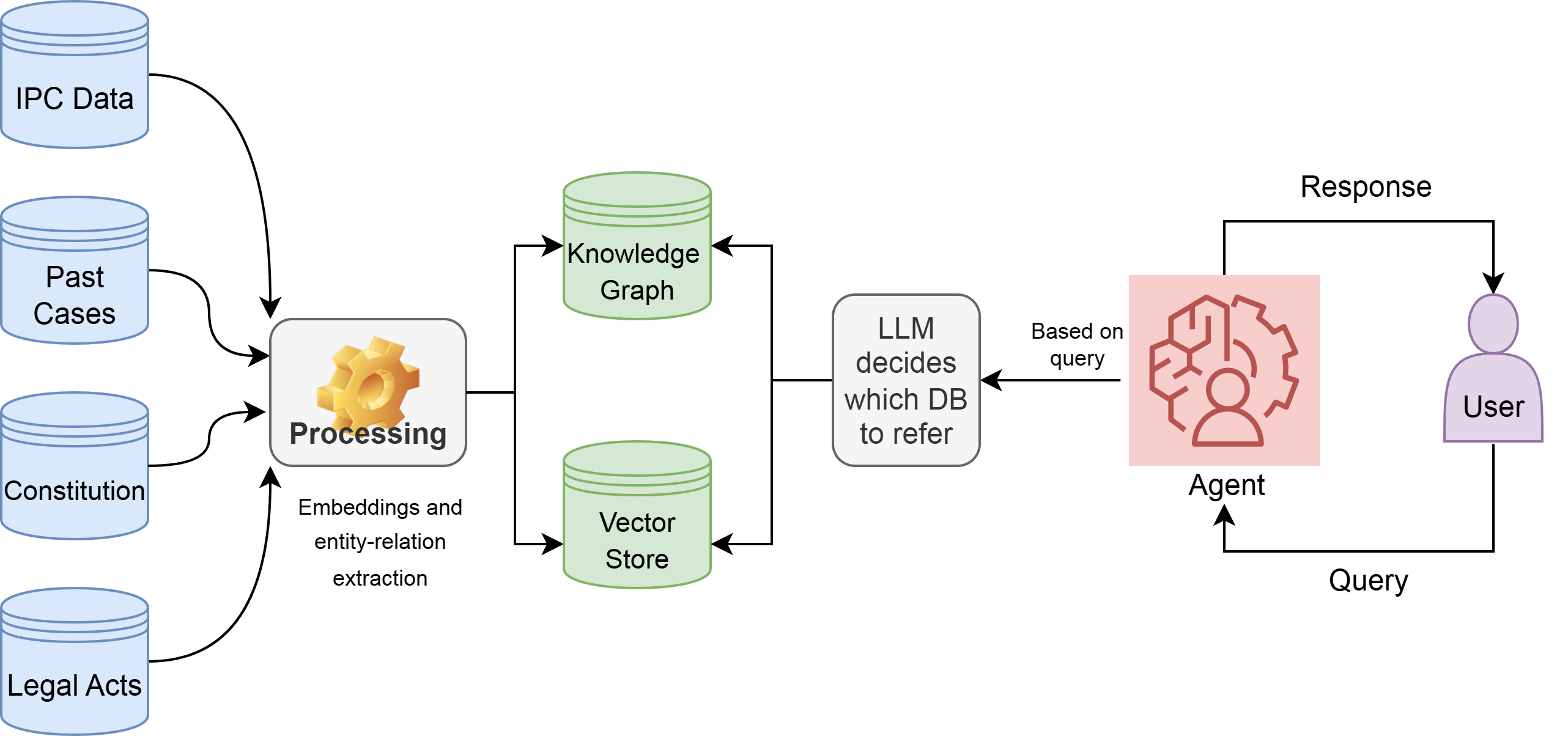}
    \caption{Hybrid architecture of the Legal AI system integrating RAG modules, Knowledge Graph, and agentic workflow.}
    \label{fig:arch_final}
\end{figure}

Figure~\ref{fig:arch_final} illustrates the hybrid architecture integrating
three domain-specific RAG pipelines with a Neo4j-based Knowledge Graph and an agentic controller.

\subsection{System Design Overview}
Indian legal documents exhibit strong domain heterogeneity: statutes and constitutional provisions are hierarchically structured, penal code sections model offense–punishment relations, and case law forms deep citation networks. Treating these materials as a single undifferentiated corpus often leads to retrieval drift and reasoning failures. To address this, we adopt a domain-partitioned retrieval strategy in which the legal corpus is divided into three RAG pipelines: the Past Cases RAG, which captures judicial reasoning, holdings, and citation structures; the Statutes and Constitution RAG, which models hierarchical statutory provisions and fundamental rights; and the IPC RAG, which represents offense definitions, punishments, and procedural classifications. Each module functions as an independent semantic retriever optimized for its respective domain, reducing cross-domain interference and improving retrieval precision for structurally diverse documents. 

Complementing these RAG modules, a large-scale Neo4j-based Legal Knowledge Graph captures relationships such as Case Case citations, Case Judge assignments, Statute Case references, and IPC offense-punishment links, enabling relational and multi-hop queries that cannot be addressed by embedding similarity alone. All components operate under an agentic orchestrator that interprets user queries, selects appropriate retrieval pathways, and fuses evidence into grounded outputs.

\subsection{Data Processing and Indexing}
Data ingestion is a critical step for ensuring retrieval precision. The documents were parsed, segmented by article headers through regular expressions, and normalized to preserve hierarchical numbering. Each article is divided into overlapping chunks of 1,000 characters (200 overlap) to maintain contextual continuity. Legal Acts are imported from structured CSVs, enriched with metadata such as act number, enactment year, and entity name, while Markdown syntax and hierarchical sections are standardized for embedding.

Case law documents which include around 2,900 Supreme Court judgments undergo metadata extraction and chunking after isolating sections like “Facts”, “Issues”, and “Held”. Similarly, IPC data from the FIR dataset (3,481 records) is serialized into compact text segments containing sections, offenses, and punishments. All processed text is encoded using the \texttt{all-MiniLM-L6-v2} model from SentenceTransformers (384-dim) and indexed in ChromaDB with an HNSW graph under cosine similarity. Each RAG maintains its own vector collection, while a hybrid retriever fuses similarity scores across collections through weighted combination, enabling domain-specific balancing between constitutional and statutory queries.

\subsection{Retrieval and Generation Pipeline}
At query time, the system computes a query embedding and retrieves the top-k (default k=5) most relevant chunks from the indexed collections. Retrieval is performed either from a single corpus or in hybrid mode, depending on the detected query type. The retrieved documents are then aggregated into a structured prompt formatted for the Gemini 2.5 Flash model, which performs controlled answer generation. The prompt template enforces grounding through explicit constraints such as: “Base your answer strictly on the provided legal documents and cite Articles, Acts, or Sections where possible.” 
This approach ensures factual traceability and mitigates hallucination risks, 
in line with best practices proposed by \citet{lawllm2024} and \citet{lexrag2025}.

The generated response is post-processed to extract referenced entities and cite source segments. The retrieval and generation subsystems are designed to be modular and stateless, enabling independent scaling and evaluation. The pipeline is implemented in Python with support for command-line and interactive query modes, providing reproducibility and ease of experimentation.

\subsection{Knowledge Graph Construction}
To address the limitations of text-only retrieval, the system incorporates a Neo4j-based legal knowledge graph that explicitly represents the interconnections among legal entities. The KG is designed to facilitate relational reasoning, linking cases to the statutes they interpret, and statutes to the offenses and punishments they prescribe. The architecture comprises four sequential stages: data ingestion, LLM-based entity extraction, graph modeling, and relational querying.

Entity and relationship extraction are performed using Google Gemini 2.0 Flash through two structured API calls per document: one for summarization and another for entity extraction. The model is prompted to identify eleven entity types, cases, courts, judges, IPC sections, articles, legal acts, legal principles, parties, offenses, punishments, and citations, returning structured JSON output. Extracted entities are transformed into Cypher statements for insertion into Neo4j, ensuring idempotent node creation through the MERGE operation.
To enable structured reasoning, we develop a comprehensive Legal Knowledge Graph that models case citation networks, judge-case associations, statutory references within cases, IPC offense–punishment relationships, and cross-domain links connecting statutes, the IPC, and case law. The KG consists of 2,586 nodes and 5,056 relationships, providing a unified representational layer for legal concepts.

\emph{Case HEARD\_IN Court}, 
\emph{Case DECIDED\_BY Judge}, 
\emph{Case GOVERNED\_BY IPCSection}, 
\emph{Case REFERS\_TO Article}, 
and \emph{Case CITES Case}.

The graph enables powerful query patterns that extend beyond traditional retrieval. 
For example, queries like “Find all cases that cite Article 14 and involve IPC Section 302” are resolved through multi-hop traversals combining statute and precedent layers. 
The KG thereby provides the foundation for future GraphRAG-style reasoning \cite{kglrag2025}, where structured relationships inform evidence retrieval before generation. 

\subsection{Agentic Workflow}
At the core of the system lies the agentic orchestrator, which performs query classification to determine whether the input pertains to statutes, the IPC, case law, or multi-domain reasoning. It then performs dynamic routing, directing the query to one or more RAG modules and the Knowledge Graph as required by the domain and complexity. 

The orchestrator supports parallel retrieval, collecting evidence from all relevant subsystems simultaneously, and it executes evidence fusion and grounded generation to synthesize information while maintaining transparency and citation integrity. This workflow is informed by emerging agentic retrieval research \cite{agenticlaw2025}, which demonstrates that multi-agent orchestration improves retrieval faithfulness in complex reasoning tasks.

\subsection{Evaluation Framework}
To assess system reliability, we employ a 40-question LLM-as-a-Judge evaluation suite covering constitutional law, criminal law, statutory interpretation, case citations, and multi-hop reasoning tasks. In addition to system-level evaluation, each subsystem is assessed independently: the three RAG modules and the Knowledge Graph are evaluated using embedding-based Precision@k, Recall@k, and F1 scores with a cosine similarity threshold of 0.7, while coherence and diversity metrics quantify redundancy and topic spread. Generation quality is measured using ROUGE-L and longest common subsequence to capture lexical and semantic alignment between generated answers and retrieved contexts. 

The system-level comparison includes the full hybrid architecture and a RAG-only variant. The hybrid system attains a 70\% pass rate, substantially outperforming the RAG-only baseline at 37.5\%, confirming the importance of structured relational reasoning.

% LaTeX Document: Evaluation / Experimental Setup
\section{Evaluation / Experimental Setup}

The evaluation aims to assess the effectiveness, reliability, and reasoning capabilities of the proposed hybrid RAG–KG system. In accordance with recent legal AI evaluation practices \citep{lexrag2025,legalbench2024}, the framework focuses on three components: (i) quality of the knowledge graph, (ii) performance on a synthetic legal Q/A dataset, and (iii) comparative benchmarking across system variants \citep{metrics2024b,metrics2024a}.

\subsection{Retrieval and Knowledge Representation Evaluation Setup}

The proposed system is evaluated through a unified experimental setup that assesses the quality of both the Legal Knowledge Graph (KG) and the embedding-based Vector Database Retrieval-Augmented Generation (RAG) pipeline. This dual evaluation framework is designed to measure the system’s capability to support accurate retrieval, multi-hop legal reasoning, and high-quality answer generation.

The final Legal Knowledge Graph consists of 2,586 nodes and 5,056 relationships, representing judicial cases, judges, constitutional articles, IPC sections, statutory references, and offense–punishment mappings. The KG is evaluated using a query-driven competency testing approach, where a curated set of relational queries is constructed across multiple legal categories, including constitutional provisions, IPC sections, and judge–case associations. These queries are specifically designed to test the KG’s ability to support multi-hop traversal and structured legal inference across interconnected legal entities. In addition to automated evaluation, selective manual validation is performed against authoritative legal sources to ensure relational correctness.

In parallel, the three separate Vector Database based RAG modules are evaluated using standard embedding-based information retrieval metrics, including Precision@k, Recall@k, and F1@k, computed using a cosine similarity threshold of 0.7. These metrics are also used for Knowledge Graph evaluation. These quantify the relevance and completeness of retrieved legal documents and context passages. To further assess retrieval behavior, coherence and diversity metrics are employed to measure contextual redundancy and topical spread within the retrieved results.

The quality of the generated responses is evaluated using ROUGE-L and Longest Common Subsequence (LCS) metrics. These measures capture both lexical overlap and semantic alignment between the generated answers and the corresponding ground-truth references, enabling a robust assessment of answer faithfulness and contextual grounding.

Together, this integrated evaluation framework provides a comprehensive and reproducible methodology for analyzing the performance of both structured knowledge retrieval via the KG and unstructured semantic retrieval via the vector database, as well as the end-to-end generation quality of the legal question answering system.

\subsection{LLM-as-a-Judge}
To systematically evaluate the system’s reasoning capabilities, we first constructed a 40-question synthetic legal question–answer dataset grounded directly in the actual legal corpus ingested by the system, rather than relying on externally generated or model-synthesized queries. The dataset was curated through a structured inspection of the underlying retrieval sources, including IPC statutes, constitutional provisions, Supreme Court judgments, and relational entities stored in the Knowledge Graph. Questions were formulated to reflect realistic legal research patterns inspired by IndicLegalQA \citep{indiclegalqa2025}, IL-TUR \citep{il_tur2024}, NYAYAANUMANA \citep{nyayaanumana2024}, AILQA \citep{ailqa2024}, and IL-PCSR \citep{il_pcsr2025}, while ensuring that every question was answerable from the system’s indexed data.

The dataset spans five core legal domains: constitutional interpretation, criminal law and IPC, case law and citation reasoning, statutory interpretation, and multi-hop cross-domain queries. For each question, ground-truth answers were manually extracted and verified from authoritative legal documents, including statutory texts and judicial decisions, with explicit source attribution indicating whether the answer should be retrievable via RAG modules or require relational reasoning from the Knowledge Graph. This manual validation step ensured legal accuracy, completeness, and consistency across questions, forming a reliable benchmark for evaluation within an LLM-as-a-Judge framework. 

Evaluation proceeds using an LLM-as-a-Judge scoring protocol, where system-generated answers are compared directly against the curated ground-truth responses. An independent judge model evaluates each answer along four dimensions: Correctness, Completeness, Relevance, and Legal Reasoning Quality, each scored on a 0–10 scale. The overall score is computed as the mean of these dimensions, with a threshold-based PASS/FAIL verdict. This evaluation design mirrors emerging standards in legal AI assessment and aligns with methodologies proposed in LexRAG \citep{lexrag2025} and LegalBench \citep{legalbench2024}, enabling fine-grained analysis of factual accuracy, evidential coverage, topical alignment, and quality of legal reasoning.

To isolate the contribution of individual architectural components, we benchmark two system configurations:
\begin{enumerate}
\item \textbf{Full Hybrid Pipeline}: RAG modules + Knowledge Graph + Orchestrator,
\item \textbf{RAG-Only Pipeline}: Knowledge Graph removed.
\end{enumerate}

This evaluation workflow—beginning with corpus-grounded dataset creation, followed by manual ground-truth verification, LLM-judged scoring, and comparative benchmarking—allows us to precisely quantify the impact of domain-partitioned retrieval, relational Knowledge Graph reasoning, and orchestrated evidence fusion. The hybrid system achieves a 70\% pass rate, substantially outperforming the RAG-only baseline at 37.5\%, confirming the critical role of structured relational reasoning in complex Indian legal research tasks

\section{Results}

\subsection{LLM as Judge Evaluation for System Variants}
The results in show a clear performance gap between the Knowledge Graph (KG) augmented system and the RAG-only baseline, with the hybrid pipeline achieving a substantially higher average overall score (6.09 vs. 4.16). This improvement reflects the system’s enhanced ability to ground responses in authoritative legal sources while performing relational reasoning across statutes, cases, and penal provisions. While the RAG-only system is able to retrieve semantically relevant passages, its lower overall score indicates difficulty in synthesizing complete and legally precise answers when queries require multi-hop reasoning, citation chaining, or structured interpretation of legal relationships. The higher pass rate achieved by the KG-enabled agent further reinforces that purely embedding-based retrieval is insufficient for complex legal queries that depend on explicit relational context.
\begin{table}[H]
\centering
\resizebox{\columnwidth}{!}{
\begin{tabular}{lc}
\hline
\textbf{System Variant} & \textbf{Average Overall Score} \\
\hline
Agent with Knowledge Graph & 6.09  \\
Agent without Knowledge Graph & 4.16 \\
\hline
\end{tabular}
}
\caption{Comparison of Average Overall Scores Across Different System Variants.}
\label{tab:overall_score_comparison}
\end{table}
\begin{table}[H]
\centering
\resizebox{\columnwidth}{!}{
\begin{tabular}{lccccc}
\hline
\textbf{System Variant} 
& \textbf{Correctness} 
& \textbf{Completeness} 
& \textbf{Relevance} 
& \textbf{Legal Reasoning}  \\
\hline
Agent with Knowledge Graph 
& 6.10 
& 4.97 
& 8.15 
& 5.35 \\
Agent without Knowledge Graph 
& 4.03 
& 2.67 
& 6.00 
& 3.95  \\
\hline
\end{tabular}
}
\caption{Detailed Comparison of Evaluation Metrics Across System Variants}
\label{tab:detailed_score_comparison}
\end{table}

The KG-augmented system consistently outperforms the non-KG variant across all metrics, with particularly pronounced gains in Completeness (4.97 vs. 2.67) and Legal Reasoning Quality (5.35 vs. 3.95). This indicates that access to structured relational knowledge enables the system to include critical procedural details, statutory conditions, and precedent-based reasoning that are frequently omitted by the RAG-only pipeline. While both systems achieve relatively strong Relevance scores, suggesting that retrieved content is generally on topic, the gap in Correctness and Legal Reasoning highlights that relevance alone does not guarantee legally sound answers. Overall, these results demonstrate that integrating a Knowledge Graph significantly improves not just retrieval fidelity, but the depth and coherence of legal reasoning, making the hybrid architecture more suitable for sensitive and complex legal research tasks.

\subsection{Performance of Individual Systems}
The reference-free evaluations show clear performance variation across the three RAG modules and the Knowledge Graph retriever. The Constitution and Acts RAG maintains high precision but lower recall, reflecting its tightly scoped statutory focus. The Case Law RAG exhibits moderate precision and higher diversity due to the narrative variability of judicial text. The IPC RAG demonstrates the strongest retrieval balance, achieving the highest overall F1 score and strong grounding metrics (ROUGE-L = 0.467, Token Overlap = 0.402), illustrating the consistency of structured, offense-based corpora. The Knowledge Graph retriever further improves semantic grounding, yielding the highest ROUGE-L and Token Overlap scores (both 0.653), confirming the value of explicit relational structure in legal reasoning workflows.

\begin{table}[H]
\centering
\begin{tabular}{lcc}
\hline
\textbf{Metric} & \textbf{k = 5} & \textbf{k = 10} \\
\hline
Precision@k & 0.880 & 0.800 \\
Recall@k & 0.140 & 0.240 \\
F1-Score@k & 0.224 & 0.321 \\
\hline
\textbf{ROUGE-L} & \multicolumn{2}{c}{0.152} \\
\textbf{Token Overlap} & \multicolumn{2}{c}{0.296} \\
\hline
\end{tabular}
\caption{Constitution and Legal Acts RAG performance across retrieval and generation metrics.}
\label{tab:constitution_rag}
\end{table}

\begin{table}[H]
\centering
\begin{tabular}{lcc}
\hline
\textbf{Metric} & \textbf{k = 5} & \textbf{k = 10} \\
\hline
Precision@k & 0.580 & 0.480 \\
Recall@k & 0.150 & 0.245 \\
F1-Score@k & 0.175 & 0.235 \\
\hline
\textbf{ROUGE-L} & \multicolumn{2}{c}{0.140} \\
\textbf{Token Overlap} & \multicolumn{2}{c}{0.285} \\
\hline
\end{tabular}
\caption{Past Cases RAG performance across retrieval and generation metrics.}
\label{tab:cases_rag}
\end{table}

\begin{table}[H]
\centering
\begin{tabular}{lcc}
\hline
\textbf{Metric} & \textbf{k = 5} & \textbf{k = 10} \\
\hline
Precision@k & 0.340 & 0.320 \\
Recall@k & 0.580 & 0.655 \\
F1-Score@k & 0.402 & 0.445 \\
\hline
\textbf{ROUGE-L} & \multicolumn{2}{c}{0.467} \\
\textbf{Token Overlap} & \multicolumn{2}{c}{0.402} \\
\hline
\end{tabular}
\caption{IPC RAG performance across retrieval and generation metrics.}
\label{tab:ipc_rag}
\end{table}

\begin{table}[H]
\centering
\begin{tabular}{lcc}
\hline
\textbf{Metric} & \textbf{k} & \textbf{Value} \\
\hline
Precision & \multicolumn{2}{c}{0.494} \\
Recall & \multicolumn{2}{c}{0.300} \\
F1-Score & \multicolumn{2}{c}{0.222} \\
\hline
\textbf{ROUGE-L} & \multicolumn{2}{c}{0.653} \\
\textbf{Token Overlap} & \multicolumn{2}{c}{0.653} \\
\hline
\end{tabular}
\caption{Knowledge Graph retriever performance across retrieval and grounding metrics.}
\label{tab:kg_rag}
\end{table}

\section{Conclusion}

This work presents a domain-partitioned hybrid RAG KG system designed specifically for the structure and reasoning needs of Indian law. By integrating three specialized RAG pipelines with a Neo4j legal knowledge graph and an LLM based agentic orchestrator, the system overcomes key limitations of monolithic retrieval including weak multi-hop reasoning, citation drift, and inconsistent grounding across statutes, IPC sections, and case law.

Evaluation through a 40 question LLM-as-a-Judge benchmark demonstrates clear gains: the hybrid system achieves a 70\% pass rate, significantly outperforming a RAG-only baseline at 37.5\%, while delivering stronger grounding and relational accuracy. These results show that combining semantic retrieval with structured legal knowledge provides a scalable and more interpretable foundation for future legal AI applications in the Indian context.

\section*{Limitations and Future Integration}

The current system operates as a prototype with limited data coverage: the Knowledge Graph contains only 2,586 nodes and 5,056 relations, and the RAG modules rely on small curated corpora. This restricted scope affects performance on queries requiring broader jurisprudence, recent amendments, or domain specific statutory interpretation. The orchestrator performs well on structured KG queries but struggles with open-ended legal reasoning and legal-analytics tasks, reflecting incomplete ingestion pipelines and non-production entity extraction.

Future work will focus on scaling and hardening the system. Planned extensions include expanding the KG to comprehensive Supreme and High Court coverage, enhancing RAG modules with complete statutory collections, and introducing automated ingestion and validation pipelines. We also aim to integrate controlled web grounding for time sensitive updates, develop a specialized legal-analytics engine for quantitative reasoning, and incorporate provenance tracking, confidence calibration, and fallback mechanisms. These improvements will help transition the system from prototype to a robust, domain-ready legal AI platform.

\section*{Acknowledgement}
The authors wish to acknowledge the use of ChatGPT in improving the presentation and grammar of the paper. The paper remains an accurate representation of the author's underlying contributions.
% bibliography (BibTeX)
\bibliography{references}  % file: references.bib (same folder)

\appendix
\section{Appendix}

\subsection{System Architecture and Implementation}
The entire system was developed in Python~3.13 and structured using modular components for independent RAG pipelines, a Neo4j-based Knowledge Graph (KG), and an orchestration layer. 
Each RAG module corresponds to a distinct legal domain---\textit{Constitution and Legal Acts}, \textit{Past Cases}, and \textit{Indian Penal Code (IPC)}---and exposes a uniform API interface for document ingestion, retrieval, and generation. 
All vector embeddings are generated using the \texttt{all-MiniLM-L6-v2} model from the \texttt{SentenceTransformers} library, producing 384-dimensional representations optimized for cosine similarity. 
Embedding storage and search are implemented through \texttt{ChromaDB} using an HNSW (Hierarchical Navigable Small World) index for fast approximate nearest-neighbor retrieval. 
The Neo4j graph is accessed using the \texttt{py2neo} connector through the Bolt protocol, while the orchestration logic is built atop \texttt{LangChain} and \texttt{asyncio} for asynchronous API calls.

\subsection{Prompt Templates and Retrieval Flow}
Each RAG module follows a two-stage pipeline: (1) evidence retrieval and (2) grounded generation via Gemini~2.5~Flash. 
Prompts are dynamically constructed using retrieved text chunks and metadata such as statute titles, section numbers, and case identifiers. 
A typical retrieval prompt is formatted as:
\begin{quote}
\textit{``Using only the retrieved legal excerpts below, answer the question concisely and cite relevant Articles or IPC Sections. Avoid speculation.''}
\end{quote}
To reduce hallucination, each prompt includes explicit grounding constraints and reference indicators (e.g., ``Article 14'' or ``Section 302''). 
Temperature is fixed at 0.1 and the maximum context window is set to 8192 tokens for deterministic behavior.

\subsection{Knowledge Graph Schema}
The Neo4j-based KG currently integrates over 2586 nodes and 5056 relationships across three domains. 
Entity types include \texttt{Case}, \texttt{Judge}, \texttt{Article}, \texttt{IPCSection}, \texttt{Offense}, and \texttt{Punishment}. 
Relationships comprise:
\begin{itemize}
    \item \texttt{HEARD\_IN} --- connects cases to courts,
    \item \texttt{DECIDED\_BY} --- links cases to judges,
    \item \texttt{GOVERNED\_BY} --- connects cases to relevant IPC sections,
    \item \texttt{REFERS\_TO} --- captures cross-citations between cases or statutes,
    \item \texttt{APPLIES\_TO} --- associates IPC sections with offenses.
\end{itemize}

\subsection{Dataset and Preprocessing Pipeline}
The system utilizes four primary datasets:
\begin{enumerate}
    \item \textbf{The Constitution of India} \cite{constitution}: parsed from PDF using \texttt{PyMuPDF}, segmented by article headers.
    \item \textbf{Central Legal Acts Dataset} \cite{centralacts}: imported from structured CSVs with metadata fields (\textit{act number, enactment year, entity name}).
    \item \textbf{Indian Penal Code Dataset} \cite{ipcdataset}: 3,481 IPC records from Kaggle, normalized into text templates of the form ``Section~X: Offense~--~Punishment''.
    \item \textbf{Supreme Court Case Records} \cite{pastcases}: 2,900+ case documents tokenized by section headers (\textit{Facts}, \textit{Issues}, \textit{Held}) using regex-based chunking.
\end{enumerate}
Markdown syntax, non-ASCII characters, and typographic inconsistencies were removed during preprocessing. 
Documents were chunked into 1,000-character segments with a 200-character overlap to preserve contextual continuity.

\subsection{Evaluation Metrics and Scripts}
Retrieval performance is measured using Precision@k, Recall@k, and F1@k computed from embedding cosine similarities.
Generation quality is assessed using ROUGE-L and Token Overlap \cite{metrics2024a}.
Each evaluation script supports batch processing of test queries across the three RAG instances. 
Evaluation parameters and configurations are version-controlled to ensure consistent runs.

\end{document}